\documentclass[aps,prl,superscriptaddress,twocolumn,twoside,a4paper,floatfix]{revtex4-1}
\usepackage{hyperref,graphicx,amsmath,amssymb,mathpazo,color}
\usepackage[normalem]{ulem}

\usepackage{pgfplots,tikz}
\usepackage[british]{babel}
\usepackage{amsfonts,epsfig}
\usepackage{hyperref}
\usepackage{latexsym}
\usepackage{epsfig}
\usepackage{amsmath}
\usepackage{amssymb}
\usepackage{amsfonts}
\usepackage{amsthm}
\usepackage{mathrsfs}
\usepackage{natbib}
\usepackage{color,verbatim,graphics}
\usepackage{psfrag}
\DeclareMathAlphabet{\mathrsfs}{U}{rsfs}{m}{n}
\DeclareMathAlphabet{\mathpzc}{OT1}{pzc}{m}{it}
\DeclareMathAlphabet{\matheus}{U}{eus}{m}{n}
\DeclareMathAlphabet{\mathbbold}{U}{bbold}{m}{n}

\newcommand{\ba}{\begin{eqnarray}}
\newcommand{\be}{\begin{equation}}
\newcommand{\ee}{\end{equation}}

\newcommand{\ea}{\end{eqnarray}}
\newcommand{\ban}{\begin{eqnarray*}}
\newcommand{\ean}{\end{eqnarray*}}
\newcommand{\tr}{\operatorname{tr}}%from Miguel

\newcommand{\ket}[1]{|#1\rangle}
\newcommand{\bra}[1]{\langle#1|}

\newcommand{\expect}[1]{\langle#1\rangle}

\begin{document}

\title{One-way Einstein-Podolsky-Rosen Steering}

\author{Joseph Bowles}
\affiliation{D\'epartement de Physique Th\'eorique, Universit\'e de Gen\`eve, 1211 Gen\`eve, Switzerland}
\author{Tam\'as V\'ertesi}
\affiliation{Institute for Nuclear Research, Hungarian Academy of Sciences, H-4001 Debrecen, P.O. Box 51, Hungary}
\author{Marco T\'ulio Quintino}
\affiliation{D\'epartement de Physique Th\'eorique, Universit\'e de Gen\`eve, 1211 Gen\`eve, Switzerland}
\author{Nicolas Brunner}
\affiliation{D\'epartement de Physique Th\'eorique, Universit\'e de Gen\`eve, 1211 Gen\`eve, Switzerland}
\affiliation{H.H. Wills Physics Laboratory, University of Bristol, Bristol, BS8 1TL, United Kingdom}

\begin{abstract}
Einstein-Podolsky-Rosen steering is a form of quantum nonlocality exhibiting an inherent asymmetry between the observers, Alice and Bob. A natural question is then whether there exist entangled states which are one-way steerable, that is, Alice can steer Bob's state, but it is impossible for Bob to steer the state of Alice. So far, such a phenomenon has been demonstrated for continuous variable systems, but with a strong restriction on allowed measurements, namely considering only Gaussian measurements. Here we present a simple class of entangled two-qubit states which are one-way steerable, considering arbitrary projective measurements. This shows that the nonlocal properties of entangled states can be fundamentally asymmetrical. \end{abstract}

\maketitle

The nonlocality of entangled quantum states, first pointed out by Einstein, Podolsky and Rosen \cite{EPR}, was later proven by Bell \cite{bell} to be an inherent feature of the theory. Nowadays quantum nonlocality is considered as a fundamental aspect of the theory, and plays a central role in quantum information processing \cite{horo,review}.

The concept of steering (or EPR-steering), proposed by Schr\"odinger \cite{schrodinger}, brings an alternative approach to this phenomenon. Consider two remote observers, Alice and Bob, who share a pair of entangled particles. By performing a measurement on her system, Alice can steer the state of Bob's system. Importantly, it is the intrinsic randomness of quantum theory that prevents this effect from leading to instantaneous signaling.
First explored in the context of continuous variable systems \cite{reid,reid_review}, quantum steering was recently formalized as an information-theoretic task by Wiseman et al. \cite{wiseman}. Steering finds applications in quantum information processing, e.g. for cryptography \cite{cyril} and randomness generation \cite{jd}. Experimental investigations have been reported \cite{saunders}, with notably a recent loophole-free experiment \cite{wittman}. Steering has also been discussed for detecting entanglement in Bose-Einstein condensates \cite{he2} and atomic ensembles \cite{he1}.

A characteristic trait of steering---distinguishing it from both entanglement and Bell nonlocality---is an asymmetry between the observers. As formalized in \cite{wiseman}, a steering test can be viewed as the distribution of entanglement from an untrusted party. Hence, in this protocol, Alice and Bob play different roles which are not interchangeable. Specifically, Alice tries to convince Bob that they share an entangled state. However, Bob does not trust Alice, and thus asks her to remotely steer the state of his particle according to a different measurement basis. Bob can then verify Alice's claim by checking a steering inequality \cite{cavalcanti}, as the violation of such an inequality implies the presence of entanglement. Conversely, if the inequality is satisfied, Bob will not be convinced that entanglement is present, since a local state strategy can in principle reproduce the observed data. Interestingly steering turns out to be a form of quantum nonlocality that is intermediate between entanglement and Bell nonlocality, in the sense that not all entangled states lead to steering, and not all steerable states violate a Bell inequality \cite{wiseman,saunders}.

A natural question, already raised in Ref. \cite{wiseman}, is then whether there exists one-way quantum steering. That is, are there entangled states such that steering can occur from Alice to Bob, but not from Bob to Alice? The properties of such states would thus be fundamentally different depending on the role of the observers. On the one hand Alice can convince Bob that the state they share is entangled. On the other hand, it is impossible for Bob to convince Alice that the state is entangled since the observed behaviour can be reproduced by a local state model. Note that such a phenomenon cannot occur for pure entangled states, which can always be brought to a symmetric form via local basis change (using the so-called Schmidt decomposition). Hence one-way quantum steering requires mixed entangled states. So far, it was shown theoretically \cite{olsen,olsen2} and experimentally \cite{vitus} that such phenomena can occur in continuous variable systems. However, these results hold only for a restricted class of measurements, namely Gaussian measurements, and there is no evidence that this asymmetry will persist for more general measurements. In fact, it is known that non-Gaussian measurements are useful in this context, as they are necessary to reveal the nonlocality of certain entangled states \cite{konrad}.

Here we present a simple class of two-qubit entangled states with one-way steering for arbitrary projective measurements. First we show that steering is not possible from Bob to Alice by constructing an explicit local hidden state model. Then we show that the state is nevertheless steerable when the roles of the parties are interchanged. Making use of techniques recently introduced in Skrzypczyk et al. \cite{paul}, we construct a steering test for demonstrating steering from Alice to Bob. The present work thus demonstrates a fundamental asymmetry in the nonlocal properties of certain entangled states.

We start by introducing the scenario and fixing notations. Consider two remote parties, Alice and Bob, sharing an entangled quantum state $\rho_{AB}$. By performing a local measurement on her particle, Alice can prepare the state of Bob's particle in different ways. In this work we will focus on the case of two-qubit states $\rho_{AB}$ and local qubit projective measurements. Consider that Alice measures the observable $\vec{x} \cdot \vec{\sigma}$ and obtains outcome $a=\pm1$; here $\vec{x}$ denotes a vector on the Bloch sphere, and $\vec{\sigma}=(\sigma_1,\sigma_2,\sigma_3)$ is the vector of Pauli matrices. Then, Bob's particle is left in the (unnormalized) state
\ba \rho_{a|\vec{x}}  = \tr_A(\rho_{AB} M_{a|\vec{x}} \otimes \mathbb{I} ) \ea
where $M_{a|\vec{x}} = (\mathbb{I}+ a \vec{x} \cdot \vec{\sigma})/2$ is the projector corresponding to outcome $a$. The set of unnormalized states $\{\rho_{a|\vec{x}} \}$, referred to as an assemblage, thus characterizes the experiment \cite{wiseman,paul,pusey}. The assemblage characterizes both the conditional probability of Alice's outcome, $p(a| \vec{x}) = \tr(\rho_{a|\vec{x}})$, and the (normalized) conditional state prepared for Bob $\hat{\rho}_{a|\vec{x}} =  \rho_{a|\vec{x}}/p(a| \vec{x})$. Note that all assemblages satisfy $\sum_a \rho_{a|\vec{x}} = \sum_a \rho_{a|\vec{x}'}$ for all measurement directions $\vec{x}$ and $\vec{x}'$, ensuring that Alice cannot signal to Bob.

In a steering test \cite{wiseman}, Alice wants to convince Bob that she can steer his state. Bob, who does not fully trust Alice, wants to verify her claim. In order to do so, he asks Alice to make a measurement in a given direction $\vec{x}$ (chosen from a given set of measurements), and then to announce her result $a$. By repeating this procedure a sufficient number of times, Bob can estimate the assemblage $\{\rho_{a|\vec{x}} \}$, e.g. via quantum state tomography. Bob's goal is now to find out whether
(i) Alice did indeed steer his state by making a measurement on an entangled state $\rho_{AB}$, or whether (ii) she cheated by using a local hidden state (LHS) strategy, in which no entanglement is involved. In this second case, Alice would prepare a single qubit state $\rho_\lambda$ and send it to Bob; here $\lambda$ represents a classical variable known to Alice, with an arbitrary distribution $\omega(\lambda)$. Upon receiving a measurement direction $\vec{x}$ from Bob, Alice announces an outcome $a$, according to a pre-determined strategy $p_\lambda(a|\vec{x})$. Hence Bob holds the state
\ba \label{LHS} \rho_{a|\vec{x}} = \sum_\lambda  \omega(\lambda) p_\lambda(a|\vec{x}) \rho_\lambda. \ea
Therefore, the problem for Bob is to determine whether the states in the assemblage $\{\rho_{a|\vec{x}} \}$ admit a decomposition of the form \eqref{LHS}. If this is the case, then Bob will not be convinced that Alice can steer his state. On the other hand, if no decomposition of the form \eqref{LHS} is possible, then Bob will be convinced that Alice did steer his state. More generally, we say that a state $\rho_{AB}$ is unsteerable from Alice to Bob, if the assemblage $\{\rho_{a|\vec{x}} \}$ admits a decomposition of the form \eqref{LHS} for all possible measurement directions $\vec{x}$. On the other hand, if there exists a set of measurement directions such that the corresponding assemblage $\{\rho_{a|\vec{x}} \}$ does not admit a decomposition of the form \eqref{LHS}, we say that $\rho_{AB}$ is steerable from Alice to Bob.

A steering test is thus clearly asymmetrical, as the roles played by Alice and Bob are different. Hence it is natural to ask whether there exist entangled states $\rho_{AB}$ that can be steered only in one direction, say from Alice to Bob but not from Bob to Alice. Here we show that such a phenomenon of \emph{one-way steering} occurs for simple two-qubit entangled states, considering arbitrary projective measurements.

Specifically, we consider states of the form
\begin{equation} \label{state}
\rho_{AB}(\alpha)=\alpha \Psi_- + \frac{1-\alpha}{5} \left( 2\ket{0}\bra{0}\otimes  \frac{\mathbb{I}}{2}  +3 \frac{\mathbb{I}}{2} \otimes\ket{1}\bra{1} \right)
\end{equation}
where $\Psi_- = \ket{\psi^{-}}\bra{\psi^{-}}$ denotes the projector on the singlet state $\ket{\psi^-}= (\ket{0,1}-\ket{1,0})/\sqrt{2}$ and $0 \leq \alpha \leq 1$. The state $\rho_{AB}(\alpha)$ is entangled for $\alpha > 1/19\left(-6 + 5\sqrt6\right)\simeq0.3288$, as can be checked via partial transposition \cite{horo}. We will see that in the range $0.4983 \lesssim \alpha \leq 1/2$, the state $\rho_{AB}(\alpha)$ is one-way steerable. The proof is divided into two parts. First we show that the state is unsteerable from Bob to Alice, by constructing a LHS model for $\rho_{AB}(1/2)$. Second, we show that steering can nevertheless occur from Alice to Bob, by showing that the assemblage resulting from 14 well chosen projective measurements on the state $\rho_{AB}(\alpha)$ with $\alpha \gtrsim 0.4983$ does not admit a decomposition of the form \eqref{LHS}.

\emph{No steering from B to A.} We construct a LHS model, from Bob to Alice, for arbitrary local projective measurements on $\rho_{AB}(1/2)$.
The model works as follows. Bob first sends to Alice a pure qubit state of the form
\ba \rho_{\lambda} = (\mathbb{I}+ \lambda_0 \vec{\lambda} \cdot \vec{\sigma})/2 \ea
where $\lambda_0= \pm 1$, and $\vec{\lambda}  = (\lambda_1,\lambda_2,\lambda_3) = (\cos{\phi}\sin{\theta},\sin{\phi}\sin{\theta},\cos{\theta})$ is a vector on the Bloch sphere distributed according to the density
\ba \omega(\theta,\phi)= \frac{1}{2 \pi} \cos^2(\theta/2) . \ea
That is, the probability of using a given vector $\vec{\lambda}$ depends on its height on the Bloch sphere. Note that $\lambda_0$ and $\vec{\lambda}$ represent here the classical variables available to Bob. Upon receiving an arbitrary measurement direction $\vec{y}= (y_1,y_2,y_3)$ from Alice, Bob then announces outcome $b= -  \lambda_0 \text{sgn}(\vec{y}\cdot \vec{\lambda})$. Finally, Alice characterizes her state. For convenience, we consider here the case where she performs an arbitrary projective measurement along direction $\vec{x}=(x_1,x_2,x_3)$ with outcome $a$.

Now we compute the statistics of the above model, focusing first on the case $\lambda_0=1$. Due to the form of the state \eqref{state}, we can take $\vec{y}= (0, \sin{\theta_B},\cos{\theta_B})$ without loss of generality.
Moreover, it will be convenient to use a new reference frame such that the $\hat{e}_3=(0,0,1)$ axis is aligned on Bob's vector $ \vec{y}$. Angles and axes in the new frame are denoted with a tilde. First we evaluate the distribution of $\vec{\lambda}$ in the new frame. That is, we compute $\omega(\tilde{\theta},\tilde{\phi})$, with $ \vec{\lambda} = (\tilde{\lambda}_1,\tilde{\lambda}_2,\tilde{\lambda}_3) = (\cos{\tilde{\phi}}\sin{\tilde{\theta}},\sin{\tilde{\phi}}\sin{\tilde{\theta}},\cos{\tilde{\theta}})$. 
Since the new frame is obtained by performing a rotation of $-\theta_B$ around the $\hat{e}_1=(1,0,0)$ axis, we have that $\lambda_3 = -\sin{\theta_B} \tilde{\lambda}_2+ \cos{\theta_B} \tilde{\lambda}_3 $.
Moreover, since $\theta = \text{arcos}(\lambda_3)$, we have that
\ba \omega(\theta,\phi) = \frac{1}{2\pi}\cos^2 \left( \frac{\text{arcos}(\lambda_3)}{2} \right)   = \frac{1+ \lambda_3 }{4\pi}  . \ea
Hence we get that
\ba \nonumber \omega(\tilde{\theta},\tilde{\phi}) = (1-\sin{\theta_B} \sin{\tilde{\phi}}\sin{\tilde{\theta}} + \cos{\theta_B} \cos{\tilde{\theta}})/4\pi .\ea
Next, we write Alice's vector in the new frame,
$\vec{x} = (\cos{\tilde{\phi}_A}\sin{\tilde{\theta}_A},\sin{\tilde{\phi}_A}\sin{\tilde{\theta}_A},\cos{\tilde{\theta}_A})$.
Using the fact that $\tr(\vec{x}\cdot \vec{\sigma} \rho_\lambda)=\vec{x}\cdot \vec{\lambda} $, we obtain the correlation
\ba  \expect{ab} &=&  -  \int_{0}^{2\pi} \text{d} \tilde{\phi}  \int_{0}^{\pi} \sin{\tilde{\theta}} \text{d}\tilde{\theta} \, \omega(\tilde{\theta},\tilde{\phi}) (\vec{x} \cdot \vec{\lambda} ) \text{sgn}(\vec{y}\cdot \vec{\lambda})   \nonumber \\
&=&    \int_{0}^{2\pi} \text{d} \tilde{\phi} \bigg( \int_{\pi/2}^{\pi} \sin{\tilde{\theta}} \text{d}\tilde{\theta} \, \omega(\tilde{\theta},\tilde{\phi})  (\vec{x} \cdot \vec{\lambda} )  \\
&  &   \quad \quad \quad   -  \int_{0}^{\pi /2} \sin{\tilde{\theta}} \text{d}\tilde{\theta} \, \omega(\tilde{\theta},\tilde{\phi})  (\vec{x} \cdot \vec{\lambda} )   \bigg)  \nonumber \ea
Since $\vec{x} \cdot \vec{\lambda} = \sin{\tilde{\theta}}\sin{\tilde{\theta}_A}\cos(\tilde{\phi}-\tilde{\phi}_A) + \cos{\tilde{\theta}}\cos{\tilde{\theta}_A}$, we find after a lengthy but straightforward calculation
\ba \label{corr} \expect{ab} = -\frac{\cos{\tilde{\theta}_A}}{2} = -\frac{\vec{x} \cdot \vec{y}}{2}. \ea
Note that $\tilde{\theta}_A$ is the angle between vectors $\vec{x}$ and $\vec{y}$.

Finally we calculate the marginals, i.e. the local expectation values for Bob
\ba \label{mB} \expect{b} &=& -  \int_{0}^{2\pi} \text{d} \tilde{\phi}  \int_{0}^{\pi} \sin{\tilde{\theta}} \text{d}\tilde{\theta} \, \omega(\tilde{\theta},\tilde{\phi}) \text{sgn}(\vec{y}\cdot \vec{\lambda})\nonumber \\
&=&   - \frac{\cos{\theta_B}}{2} = - \frac{y_3}{2} \ea
and for Alice
\ba  \label{mA} \expect{a} &=&   \int_{0}^{2\pi} \text{d} \phi  \int_{0}^{\pi} \sin{\theta} \text{d}\theta \, \omega(\theta,\phi) (\vec{x}\cdot \vec{\lambda})  \nonumber \\
&=&   \frac{\cos{\theta_A}}{3} =  \frac{x_3}{3}. \ea
Note that for computing Alice's marginal, it is more convenient to use the original reference frame.

At this point, it is useful to note that the correlations \eqref{corr} correspond exactly to those of the state $\rho_{AB}(1/2)$. Moreover, the marginals \eqref{mB} and \eqref{mA} have the right form, but are in fact slightly stronger than those of $\rho_{AB}(1/2)$. In order to weaken the marginals, while keeping the correlation unchanged, we now use the variable $\lambda_0$. Specifically, consider the distribution $p(\lambda_0=-1)=f$. Hence the marginals are decreased to
$\expect{a} = (1-2f)  x_3/3$ and $\expect{b} = (1-2f)  y_3/2$. Choosing a flipping probability of $f=1/5$, we finally get
\ba \expect{a} = \frac{ x_3 }{5}  \, , \,\,  \expect{b} = \frac{3 y_3 }{10}  \, , \,\,   \expect{ab} = -\frac{\vec{x} \cdot \vec{y}}{2} . \ea
Hence the model simulates exactly the statistics of local projective measurements on the state $\rho_{AB}(1/2)$.
The assemblage $\{\rho_{b|\vec{y}} \}$ observed by Alice is thus identical to the assemblage expected for the state $\rho_{AB}(1/2)$, that is, $\rho_{b|\vec{y}}  = \tr_B(\rho_{AB}(1/2)  \mathbb{I} \otimes  M_{b|\vec{y}}  )$, where $M_{b|\vec{y}}=(\mathbb{I}+b\vec{y} \cdot \vec{\sigma})/2$. Therefore the state $\rho_{AB}(1/2)$ is unsteerable from Bob to Alice. The extension to the case $\alpha<1/2$ is straightforward.

Finally, note that the above model can also be understood as a local hidden variable model (LHV); thus the statistics of local projective measurements on $\rho_{AB}(\alpha)$ with $\alpha \leq 1/2$ cannot violate any Bell inequality \cite{footnote1}. This complements a series of works describing entangled states admitting a LHV model \cite{werner,barrett,mafalda,flavien}.

\emph{Steering from A to B.} We will see now that the situation is completely different when the roles of Alice and Bob are interchanged. Specifically, the state $\rho_{AB}(\alpha)$ with $\alpha \gtrsim 0.4983$ is steerable from Alice to Bob. In order to prove this, we will show that, for a well chosen set of $m$ projective measurements for Alice, the resulting assemblage $\{\rho_{a|\vec{x}} \}$ obtained on Bob's side cannot be reproduced by any LHS model.

The observables measured by Alice are denoted $A_i = \vec{x}_i \cdot \vec{\sigma}$ with $i=1,...,m$ and outcome $a=\pm1$. Bob characterizes the state $\rho_{a|\vec{x}_i}$ by tomography, making measurements represented by the Pauli matrices, $\sigma_j$ with $j=1,2,3$, outcome $b=\pm1$, and $\sigma_0= \mathbb{I}$. The observed statistics are then given by
\ba \label{stat} \expect{ab}_{ij} &=& \tr( \rho_{AB}(\alpha) A_i \otimes \sigma_j)   \\ \nonumber
\expect{b}_j &=& \tr( \rho_{AB}(\alpha)  \mathbb{I} \otimes \sigma_j).  \ea
Alice's marginals are given by $ \expect{a}_i = \expect{ab}_{i0}$.

Considering a given number of measurements $m$, we now aim at finding the largest value of $\alpha$, denoted $\alpha^*$, for which the state $\rho_{AB}(\alpha)$ is unsteerable from Alice to Bob. That is, we want to determine the largest $\alpha$ such that the statistics \eqref{stat} can be reproduced by a LHS model, i.e.
\ba \expect{ab}_{ij} = \sum_\lambda E_\lambda(i)  \tr(\rho_\lambda \sigma_j )  \quad
\expect{b}_{j} = \sum_\lambda   \tr(\rho_\lambda \sigma_j ) \ea
where $E_\lambda(i)= p_\lambda(a=1|i)-p_\lambda(a=-1|i)$ is the expectation value of Alice's outcome $a$ for a given $\lambda$ and measurement $i$. Note that here the local states $\rho_\lambda$ are not normalized, and one has that $\sum_\lambda \tr(\rho_\lambda) = 1$.

To solve this problem we make use of a semi-definite programming (SDP) technique recently developed in \cite{paul}, for deciding whether a given assemblage $\{\rho_{a|\vec{x}} \}$ belongs to the set of 'unsteerable assemblages', that is, whether $\{\rho_{a|\vec{x}} \}$ admits a decomposition of the form \eqref{LHS}. Our present problem can be solved by the following SDP:
\begin{equation} \label{SDP}
\begin{aligned}
&\alpha^* \equiv \max{  \alpha} \\
\text{s.t.}\, &   \sum_\lambda E_\lambda(i)\tr\left(\rho_\lambda \sigma_j \right)=\expect{ab}_{ij} , \quad   \sum_\lambda   \tr(\rho_\lambda \sigma_j )=\expect{b}_{j} \\
&  \tr \sum_\lambda \rho_\lambda = 1, \quad \rho_\lambda \geq 0 \quad\forall \lambda,
\end{aligned}
\end{equation}
where the optimization variables are $\rho_\lambda$, and the quantities $\expect{ab}_{ij}$ and $\expect{b}_{j} $ are computed as in \eqref{stat}. Note that we can focus here on LHS strategies for which Alice provides a deterministic outcome $a$ given $\lambda$ and $i$ \cite{paul}, that is $E_\lambda(i)=\pm1$ for $i=1...m$. Hence we have altogether $2^m$ possible strategies for Alice to consider. The above SDP is then implemented for each strategy.

Using the above SDP, we can thus estimate, for a particular choice of $m$ measurement directions $ \vec{x}_i $ (with $i=1...m$), the threshold value $\alpha^*$ for which the state $\rho_{AB}(\alpha)$ is steerable from Alice to Bob. For fixed $m$, we then minimize $\alpha^*$ over all possible choices of measurement operators for Alice, using a hill-climbing heuristic algorithm. Results for $m$ up to 14 are presented in Table~I. Notably, for $mÊ= 14$ we get $\alpha^* \simeq 0.4983$, thus implying that the state $\rho_{AB}(\alpha)$ with $\alpha \gtrsim 0.4983$ is steerable from Alice to Bob.

Finally, from the result of the above optimization procedure, it is in fact possible to extract an explicit steering inequality. Once the optimal measurement directions $ \vec{x}_i $ ($i=1...m$) have been found via the hill-climbing algorithm, the dual of the SDP problem \eqref{SDP} allows us to extract a linear steering inequality \cite{paul} of the form
\ba \sum_{i=1}^m \sum_{j=1}^3 s_{ij}  \expect{ab}_{ij}  +  \sum_{i=1}^m  s^A_{i}  \expect{a}_{i}  + \sum_{j=1}^3 s^B_{j}  \expect{b}_{j} \leq L . \ea
Such an inequality is characterized by (i) a set of real coefficients: $s_{ij}$, $s^A_{i}$, and $s^B_{j}$, and (ii) a bound $L$ which holds for any LHS strategy. In the supplementary material, we follow the above method to give explicitly a steering inequality featuring $m=13$ measurements, which is violated by performing appropriate measurements (which we give as well) on the state $\rho_{AB}(1/2)$.

 \begin{table}[t!]
\caption{Threshold values $\alpha^*$ for which the state $\rho_{AB}(\alpha)$ is steerable from Alice to Bob. The optimization is conducted over all possible steering tests where Alice performs $m=2,\ldots,14$ projective measurements.}
\centering
\begin{tabular}{c|ccccccc}
\hline
$m$&2&3&4&5&6&7&8\\
  \hline\hline
  $\alpha^*$ & 0.6951 & 0.5661& 0.5424& 0.5302& 0.5156& 0.5120& 0.5088\\
$m$&9&10&11&12&13&14&$ $\\
\hline\hline
$\alpha^*$ & 0.5037 & 0.5030& 0.5014& 0.5005& 0.4993& 0.4983& $ $\\
  \hline
\end{tabular}
\end{table}

\emph{Discussion.} We have shown the existence of entangled states which are one-way steerable when considering arbitrary projective measurements. That is, the nonlocal properties of such states depend on the role played by the parties: while Alice can steer the state of Bob, it is impossible for Bob to steer Alice's state. This shows that quantum nonlocality can be fundamentally asymmetrical. An interesting open question is whether the present result can be extended to the most general measurements, i.e. POVMs. Moreover, it would be interesting to find an application, e.g. in quantum information processing, of the phenomenon of one-way steering.

\emph{Acknowledgements.} We thank Jonatan Brask for discussions, and acknowledge financial support from the Swiss National Science Foundation (grant PP00P2\_138917 and QSIT), SEFRI (COST action MP1006) and the EU DIQIP. T.V. thanks support from the J\'anos Bolyai Programme, the OTKA (PD101461) and the T\'AMOP-4.2.2.C-11/1/KONV-2012-0001 project.

\section{Appendix}

Here we describe explicitly the steering test witnessing the fact that the state $\rho_{AB}(\alpha)$ (see eq. \eqref{state}) with $\alpha > (2268/2269)(1/2)$ is steerable from Alice to Bob.

Here we consider the case of $m=13$ measurement settings for Alice, characterized by operators of the form $A_i = \vec{x}_i \cdot \vec{\sigma}$ with $i=1,...,m$ with outcome $a=\pm1$. Bob performs tomography, making measurements in the Pauli basis $\sigma_j$ with $j=1,2,3$, outcome $b=\pm1$. The observed data is then given by
\ba \expect{ab}_{ij} &=& \tr( \rho_{AB}(\alpha) A_i \otimes \sigma_j) \\
\expect{b}_j &=& \tr( \rho_{AB}(\alpha)  \mathbb{I} \otimes \sigma_j) \\
\expect{a}_{i} &=& \tr( \rho_{AB}(\alpha) A_i \otimes \mathbb{I}). \ea

We now construct a linear steering inequality of the form
\ba \label{steering ineq} \sum_{i=1}^m \sum_{j=1}^3 s_{ij}  \expect{ab}_{ij} + \sum_{i=1}^m  s^A_{i}  \expect{a}_{i}  +  \sum_{j=1}^3 s^B_{j}  \expect{b}_{j} \leq L . \ea
The inequality is characterized by the matrix {\bf S}, with real coefficients $s_{ij} $, and the vectors ${\bf S^A}$ and ${\bf S^B}$, with real elements $s^A_{i} $ and $s^B_{j} $, respectively. Specifically, we have that

\begin{equation}\label{ineq}
{\bf S} = \left(
\begin{tabular}{ccc}
$\frac{1}{12}$ & $\frac{6}{175}$ & $\frac{-12}{217}$ \\
$\frac{-9}{79}$ &	$\frac{-1}{38}$ &	$\frac{-7}{94}$ \\
$\frac{-1}{162}$ &	$\frac{18}{133}$ &	$\frac{-1}{18}$ \\
$\frac{17}{157}$ &	$\frac{-6}{143}$	& $\frac{-10}{141}$ \\
$\frac{5}{62}$ &	$\frac{-10}{97}$ &	$\frac{-1}{62}$ \\
0 &	$\frac{2}{103}$ &	$\frac{-9}{76}$ \\
$\frac{-16}{105}$ &	$\frac{1}{89}$ &	0 \\
$\frac{5}{104}$ &	$\frac{-6}{79}$ &	$\frac{-11}{72}$ \\
$\frac{-4}{73}$ &	$\frac{-6}{109}$ &	$\frac{-54}{433}$ \\
$\frac{-3}{26}$ &	$\frac{-3}{20}$ &	$\frac{-2}{83}$ \\
$\frac{10}{179}$	& $\frac{9}{103}$	& $\frac{-13}{121}$ \\
$\frac{1}{132}$ &	$\frac{-4}{33}$ &	$\frac{-2}{49}$ \\
$\frac{-11}{107}$ &	$\frac{14}{139}$ &	$\frac{-20}{161}$
\end{tabular}
\right),\,\,
{\bf S^A} = -\left(
\begin{tabular}{c}
$\frac{1}{71}$ \\
$\frac{1}{53}$ \\
$\frac{1}{71}$ \\
$\frac{2}{111}$ \\
$\frac{1}{244}$ \\
$\frac{3}{100}$ \\
0 \\
$\frac{4}{103}$ \\
$\frac{2}{63}$ \\
$\frac{1}{163}$ \\
$\frac{3}{110}$ \\
$\frac{1}{96}$ \\
$\frac{3}{95}$
\end{tabular}
\right), \,\,
{\bf S^B} = \left(
\begin{tabular}{c}
$0$ \\
$0$ \\
$\frac{-15}{59}$
\end{tabular} \right).
\end{equation}
The local bound of the above inequality, which holds for any possible LHS model, is $L=1$. This can be verified using e.g. the techniques of Refs \cite{saunders,pusey}.

Now we give the measurement operators for Alice, characterized by Bloch vectors $\vec{x}_i $ with $i=1,...,13$. We have that

\begin{equation}\label{eq16}
{\bf V} = \left(
\begin{tabular}{ccc}
$\frac{-31}{38}$ & $\frac{-17}{54}$ & $z_1$ \\
$\frac{69}{82}$ &	$\frac{6}{35}$ &	$z_2$ \\
$\frac{5}{111}$ &	$\frac{-103}{110}$ &	$z_3$ \\
$\frac{-9}{11}$ &	$\frac{7}{23}$	& $z_4$ \\
$\frac{-52}{83}$ &	$\frac{53}{69}$ &	$z_5$ \\
$\frac{-1}{673}$ &	$\frac{-8}{49}$ &	$z_6$ \\
$\frac{456}{457}$ &	$\frac{-5}{83}$ &	$z_7$ \\
$\frac{-128}{427}$ &	$\frac{57}{124}$ &	$z_8$ \\
$\frac{37}{90}$ &	$\frac{35}{88}$ &	$z_9$ \\
$\frac{47}{77}$ &	$\frac{183}{233}$ &	$z_{10}$ \\
$\frac{-37}{94}$	& $\frac{-53}{86}$	& $z_{11}$ \\
$\frac{-3}{58}$ &	$\frac{116}{121}$ &	$z_{12}$ \\
$\frac{13}{23}$ &	$\frac{-76}{137}$ &	$z_{13}$
\end{tabular}
\right),
\end{equation}
where the $k$-th row of the above matrix is understood to be $\vec{x}_k $. By normalization of the vectors, we have that $z_k^2 =1 - v_{k1}^2 - v_{k2}^2$ where $v_{ij}$ denote the elements of matrix {\bf V}, and $z_k$ is chosen to be positive. With this set of measurements performed on the state $\rho_{AB}(1/2)$, we can evaluate the right hand side of \eqref{steering ineq} giving us the value
$S_q > \frac{2269}{2268} > 1 = L$ hence demonstrating steering from Alice to Bob.


\begin{thebibliography}{99}
\bibitem{footnote1} Note that our local model is related to that of Ref. \cite{werner}; the central difference is that we consider here a non-uniform distribution of vectors $\vec{\lambda}$.
\end{thebibliography}
\end{document}